\documentclass[aps,prl,twocolumn,groupedaddress]{revtex4}%

\usepackage{graphicx}
\usepackage{delarray}
\usepackage{amsmath}
\usepackage{amsfonts}
\usepackage{amssymb}%
\begin{document}
\title{Comment on "Universally Diverging Gruneisen Parameter and the
Magnetocaloric Effect Close to a Quantum Critical Point".}
\author{Mucio A. \surname{Continentino} }
\affiliation{Instituto de F\'{\i}sica - Universidade Federal Fluminense\\
Av. Litor\^anea s/n,  Niter\'oi, 24210-340, RJ - Brazil}


\maketitle

In a recent Letter Zhu et al. \cite{zhu} obtained scaling relations
for the Gr\"{u}neisen parameter $\Gamma\propto \beta/C$, where
$\beta$ is the thermal expansion and C the specific heat, of a heavy
fermion system at the quantum critical point (QCP). We show here
that generally this quantity yields information on the
\textit{shift} exponent governing the critical line of finite
temperature transitions and not on the crossover exponent $\nu z$ as
obtained in Ref. \cite{zhu}. We start with the Ehrenfest equation
\cite{samara} relating the pressure derivative of the line of
critical temperatures, $T_{N}$, to these thermodynamic quantities,
\begin{equation}
\frac{dT_{N}}{dP}=VT\frac{\Delta\beta}{\Delta C}=V\frac{\Delta\beta}{\Delta
C/T}\label{ehrenfest}%
\end{equation}
where $\Delta\beta$ and $\Delta C$ are {\em the differences in
thermal expansion and specific heat in the two phases} (the
critical part), respectively. The thermal expansion is defined by
\begin{equation}
\beta=\frac{1}{V}\frac{\partial V}{\partial T}=-\kappa_{T}\frac{\partial^{2}%
F}{\partial T\partial V}|_{N}.
\end{equation}
The isothermal compressibility $\kappa_{T}=-(1/V)({\partial
V}/{\partial P})$ is non-singular at the QCP. The volume thermal
expansion, $\beta=2\alpha_{a}+\alpha_{c}$ where $\alpha_{i}$ are
the linear thermal expansion coefficients along different axes
$i$. The critical line of finite temperature phase transitions
close to the QCP is given by,
\begin{equation}
T_{N}\propto\frac{1}{u}|V-V_{c}|^{\psi}\propto\frac{1}{u}|P-P_{c}|^{\psi
}\label{tn}%
\end{equation}
which defines the shift exponent $\psi$ \cite{mucio}. The critical temperature
is reduced changing the volume by applying pressure in the
system. The quantities $V_{c}$ and $P_{c}$ are the critical volume and
pressure, respectively. From the equation above we obtain,
\begin{equation}
\frac{dT_{N}}{dP}\propto \frac{1}{u}|P-P_{c}|^{\psi-1}\propto T_{N}^{1-\frac{1}{\psi}%
},\label{deri}%
\end{equation}
such that,

\begin{equation}
\lim_{T_N \rightarrow 0} V\frac{\Delta\beta}{\Delta C/T}=
T_{N}^{1-\frac{1}{\psi}} \label{ratio}
\end{equation}
This is independent of any scaling ansatz relying only on the
shape of the critical line and the Ehrenfest equation. This result
is inconsistent with that obtained by Zhu et al.~\cite{zhu} for
$V=V_c$ and $d+z  > 4$,

\begin{equation}
\frac{\Delta\beta}{\Delta C/T} \propto T^{1-\frac{{}^{1}}{\nu z}
}.\label{qi}%
\end{equation}

The reason for the discrepancy between Ehrenfest relation's result
and the equation above, where $\psi$ is replaced by $\nu z$, is
that the purely Gaussian part of the quantum free energy which
yields the results of Zhu et al.~\cite{zhu} ignores the critical
N\'eel line. In the spin density wave theory, for $d+z \ge 4 $,
this line appears due to the dangerous irrelevant quartic
interaction $u$, as in Eqs.~\ref{tn}, with
$\psi=z/(d+z-2)$~\cite{millis}. The purely Gaussian quantum free
energy with $u=0$  is unaware of the transition  at $T_N$. These
Gaussian quantum fluctuations must then be seen as \emph{regular}
contributions with respect to the {\em thermal phase transition}
not contributing to the critical part of the physical quantities
appearing in Eq.~\ref{ehrenfest}. Our results are obtained
approaching the QCP from the magnetic side, i.e., with $T_N
\rightarrow 0$. The quantum critical point is of course a special
point of the critical line $T_N(P)$. They are particularly
relevant for systems where $T_N$ is vanishingly small but finite.

The results above are specially important since the temperature
dependence of Eq.~\ref{ratio} becomes more singular than that of
Ref.~\cite{zhu}, Eq.~\ref{qi}, with $\psi$ replaced by $\nu z$. In
the spin density wave theory, this occurs exactly for $d+z>4$,
i.e., above the upper critical dimension, $d_c=4$, which is the
condition for breakdown of hyperscaling. In this case, the
hyperscaling relation $\psi=\nu z$, which identifies the shift
with the crossover exponent, breaks down for $d+z>4$ due to the
dangerous irrelevant quartic interaction $u$ \cite{millis}.

\end{document}